\documentclass{aa}
\usepackage{graphicx}

\usepackage{txfonts}
\usepackage{natbib}
\begin{document}
   \title{Mass transfer variation in the outburst model of dwarf
   novae and soft X-ray transients}

   \author{M. Viallet \and J.-M. Hameury
          }

   \offprints{M. Viallet}

   \institute{Observatoire Astronomique, ULP \& CNRS,
              11 rue de l'Universit\'e, 67000 Strasbourg, France\\
              \email{viallet@astro.u-strasbg.fr\\}
             }

   \date{Received ; accepted }


  \abstract
   {In the standard formulation of the disc instability model for soft X-ray
   transient and dwarf nova outbursts, the mass transfer rate from the secondary
   is assumed to be constant. This may seem natural since the $L_1$ point remains
   shielded from the accretion luminosity by the accretion disc. However an
   indirect heating could take place and could lead to an enhancement of the
   mass transfer rate. It is still debated whether such an enhancement of the
   mass transfer rate during outbursts is a missing ingredient of the current model.}
   {We discuss two mechanisms that could result in an enhancement of the mass
   transfer rate during outbursts: the hot outer disc
   rim itself could significantly heat the $L_1$ point; scattered radiation by
   optically thin outflowing matter could also heat $L_1$ significantly.
   We determine quantitatively the increase of the mass transfer rate resulting
   from an extra heating.}
   {We estimate the heating of the $L_1$ point by the disc rim by analytical and
   numerical arguments. The fraction of the luminosity scattered by matter above
   the disc, here modeled as a spherically symmetric outflow, is
   determined analytically. We finally solve in a numerical way the
   vertical structure equations for the secondary star and calculate the mass transfer
   enhancement.}
   {During outbursts, the temperature at the outer edge of the disc 
   reaches $10^4$ K. The disc edge heats up the upper layer of the
secondary with a flux of the order of the intrinsic stellar flux. This probably has 
no large effect on the mass transfer rate. In soft X-ray transients, the environing medium of the disc (corona+wind) could back-scatter a certain fraction of the accretion luminosity toward $L_1$. In dwarf novae, the same effect could be due to the wind present during outburst. Since soft X-ray transients reach high luminosities, even a low efficiency of this effect could yield a significant heating of $L_1$, whereas we show that in dwarf novae this effect is negligible. Initially the incoming radiation does not penetrate below the
photosphere of the secondary. Depending on the heating efficiency, which has to be 
determined, the mass transfer rate could be significantly increased.}
{}


   \maketitle

\section{Introduction}

Dwarf novae (DNe) and Soft X-ray transients (SXTs) are semi-detached
binary systems that undergo regular outbursts. SXTs in outbursts
reach luminosities of order of $10^{38}$ erg.s$^{-1}$ in the soft
X-ray band, whereas they are faint X-ray sources during quiescence
(with luminosities $\lesssim 10^{33}$ erg.s$^{-1}$). Their outbursts
last for a few months and have a recurrence time in the range 1 --
50 years \citep[see][]{1997xrb..book.....L}. DNe are $1-5$ magnitude
brighter during outbursts than in quiescence. Their outbursts last
for a few days and recur every few weeks \citep[see][]{2003cvs..book.....W}. 
In both types of systems, a compact object (i.e. a white
dwarf in DNe and a neutrons star/black hole in SXTs) accretes from a
low mass secondary star via an accretion disc. It is now believed
that DNe and SXTs outbursts are both due to a thermal/viscous
instability of the accretion disc triggered when hydrogen becomes
partially ionized \citep[see][]{1996PASP..108...39O,2001NewAR..45..449L}. 
In this picture, the accretion disc performs a limit cycle between a cold,
quiescent state of low accretion rate and a hot, viscous state of
high accretion rate corresponding to the outburst. The difference
between DNe and SXTs timescales is thought to be due both to the
irradiation of the disc by the central source in SXTs \citep[see][]{1998MNRAS.293L..42K,2001A&A...373..251D}, and to a higher primary mass.

\begin{table*}[t]
\caption{Physical parameters for the four models condidered in this
paper. $P_{\rm hr}$ is the orbital period in hours, $a$ the orbital
separation, $R_2$ the secondary radius, $M_1$ and $M_2$ the primary
and secondary masses, $T_\star$ the effective temperature of the
secondary, $\dot{M}_{\rm tr}$ the average mass transfer rate from th
esecondary, and $\dot{M}_{\rm acc,max}$ is the maximum accretion rate onto the 
compact object. $T_\star$ for model $1-3$ is taken from \cite{2004AcA....54..181S}
and from \cite{1986A&A...162...71H} for model $4$.}            
\label{params_phys}      
\centering 
\begin{tabular}{c c c c c c c c c c c}        
\hline 
Model & System & $\mathrm{P_{orb}}$ (h) & $a\ (R_\odot)$  & $R_2/a$&
$q=M_2/M_1$ & $M_2(M_\odot)$ & $T_\star$ (K) & $\dot M_\mathrm{tr}$ (10$^{16}$
g.s$^{-1}$) & $\dot M_\mathrm{acc,max}$ (10$^{16}$ g.s$^{-1}$) & $h/r$\\  
\hline 
1 & OY Car & 1.51 & 0.6 & 0.21 & 0.1 & 0.085 &2500 & 0.5 & 16&0.06\\
2 & U Gem & 4.24 & 1.48 & 0.29 &0.36 &  0.7& 3500 & 5,10 & 100&0.09\\
3 & Z Cam  & 6.96 & 2.17 & 0.35 &0.6 &0.7 &4200& 30,60  & 250&0.1\\
\hline
4 & A 0620  & 7.8 & 3.9 & 0.21 & 0.1 & 0.7 & 3000 & 0.5 & $10^3$&0.1\\
\hline 
\end{tabular}
\end{table*}

The thermal/viscous instability alone is not sufficient to explain
the rich and diverse features of DNs and SXTs lightcurves. For
example, \citet{1989PASJ...41.1005O} suggested that a tidal instability driving the
disc eccentric \citep[see][]{1988MNRAS.232...35W} would account for the
superoutburst phenomenon of SU UMa stars (a subclass of DNe) when
coupled to the standard thermal/viscous instability. Tidal effects
have also been proposed to account for the secondary maximum often
present in SXTs lightcurves \citep[see][]{2002MNRAS.337.1329T}.

In its most common formulation, the disc instability model (DIM)
assumes that the mass transfer rate from the secondary is fixed, and
given by the secular losses of angular momentum from the binary
system. This assumption is most likely to be wrong and significant
deviation of the mass transfer rate from its secular mean on time
scales as short as the outburst duration could be an essential
ingredient missing in the standard model. Numerical investigations
showed that variations of the mass transfer rate could explain the
outburst bimodality of DNe \citep{1999AcA....49..383S} and that it could lead to
long outbursts similar to superoutbursts \citep{2000NewAR..44...15H}. 
\cite{1993A&A...279L..13A} and \cite{1993ApJ...408L...5C} sought to explain the
rebrightenings of SXTs lightcurves by an increase of the mass
transfer rate. However, in these works either an empirical
relationship between the mass transfer rate and the mass accretion
rate was assumed or mass transfer burst episodes were imposed
without physical grounds. It has been often argued that a such mass
transfer enhancement could result from irradiation of the secondary
during an outburst. \cite{2007A&A...475..597V} investigated
this problem and found that, because the $L_1$ point is shielded
from direct irradiation disc by the secondary star, the increase of
the mass transfer rate is modest.

In this paper, we investigate two different mechanisms that could be
responsible for a mass transfer enhancement, resulting in both cases
from an increase of the temperature at the $L_1$ point. Since $L_1$
lies in the shadow cast by the accretion disc, any heating must be
indirect. The first effect that we consider is heating by radiation
emitted by the hot edge of the disc during an outburst, which
produces a flux at $L_1$ comparable to the intrinsic stellar flux.
The second effect that we consider is the heating of $L_1$ by the
accretion luminosity scattered by a wind and/or a corona extending
above the accretion disc. The scattered fraction is small, but the
scattered flux is still significant in SXTs due to their high luminosity
in outburst. 
We then turn to the determination of the mass transfer enhancement.
First, we compute the vertical structure of the secondary
envelope in DNe and SXTs both in quiescence and outbursts to
determine the dependence of the mass transfer rate versus a given
incoming heating flux. Our results show that $\dot{M}$ can potentially 
be raised by a factor of up to $\sim 100$ in SXTs, and up to $\sim 10$ in DNe.
We then discuss if the effects investigated here could lead to an efficient 
heating of the $L_1$ point. This is done by computing the fraction of the 
incoming radiation that penetrates the photosphere. This depends on 
the nature of the radiation, hence on its origin. We show that the thermal flux 
emitted by the edge of the disc can penetrate below the photosphere of the secondary. 
The resulting heating is probably too small to have any major effect. For soft X-ray transients, the 
backscattered radiation is in the soft X-ray band and is therefore strongly absorbed by the initially neutral hydrogen in the very upper part of the atmosphere. However, this large X-ray flux significantly affects the atmosphere. A full analysis of the radiative transfer should be undertaken to determine the exact structure of the atmosphere.


The paper is organized as follow: in Section 2, we discuss the two
mechanisms mentioned above. Section 3 is devoted to the computation
of the vertical structure of secondary stars in DNe and SXTs, and to
the relationship between the mass transfer enhancement and the
heating of $L_1$. Concluding remarks are given in section 4.

\section{Heating flux at the $L_1$ point during outburst}

In the following, we considere 4 models, with parameters given in
Table 1, that are representative of various DNe subclasses and SXT outbursts. Models 1-3 correspond to dwarf novae with increasing
orbital periods, and model 4 corresponds to A 0620-00, emblematic of
short period soft X-ray transients.

\subsection{Heating by the disc rim}

During an outburst, the disc grows as a consequence of an effective
outward transport of angular momentum. The maximum disc size is
determined by tidal interactions with the secondary which
effectively truncate the disc at a mean radius $R_\mathrm{tid} \sim
0.85 R_{L_1}$ \citep[see][]{2002AcA....52..263S,1994PASJ...46..621I}. During an
outburst, the disc will reach this maximal radius, for SU UMa
systems this would be true only during superoutbursts 
\cite[see][]{1989PASJ...41.1005O,1996PASP..108...39O}, 
 and tidal dissipation heats up the outer region of the
disc. \cite{2002AcA....52..263S} argued that tidal dissipation occurs in
a very narrow region at the disc rim and suggested that this energy
is radiated away by the disc edge, whereas heat generated by viscous
dissipation is radiated by the surface of the disc; this contrasts
with our previous work, where tidal dissipation was considered to
occur on a relatively large area of the disc, significantly larger
than the disc thickness \citep[see][]{2001A&A...366..612B, 2005A&A...443..283H}. 
We do not discuss this issue further here and we use Smak's assumption which maximizes the energy radiated by the disc edge. We show below that, even in this very favorable hypothesis, the effect is small.

In this picture, removal of angular momentum by the tidal torque is
negligible inside the truncation radius. The only effect of the
tidal torque is to prevent the disc from exceeding the radius
$R_\mathrm{tid}$. In this prescription for tidal effects, the only
free parameter is $R_\mathrm{tid}$.

\begin{figure*}[t]
\parbox{0.5\linewidth}{\center \includegraphics[width=6cm,angle=90]{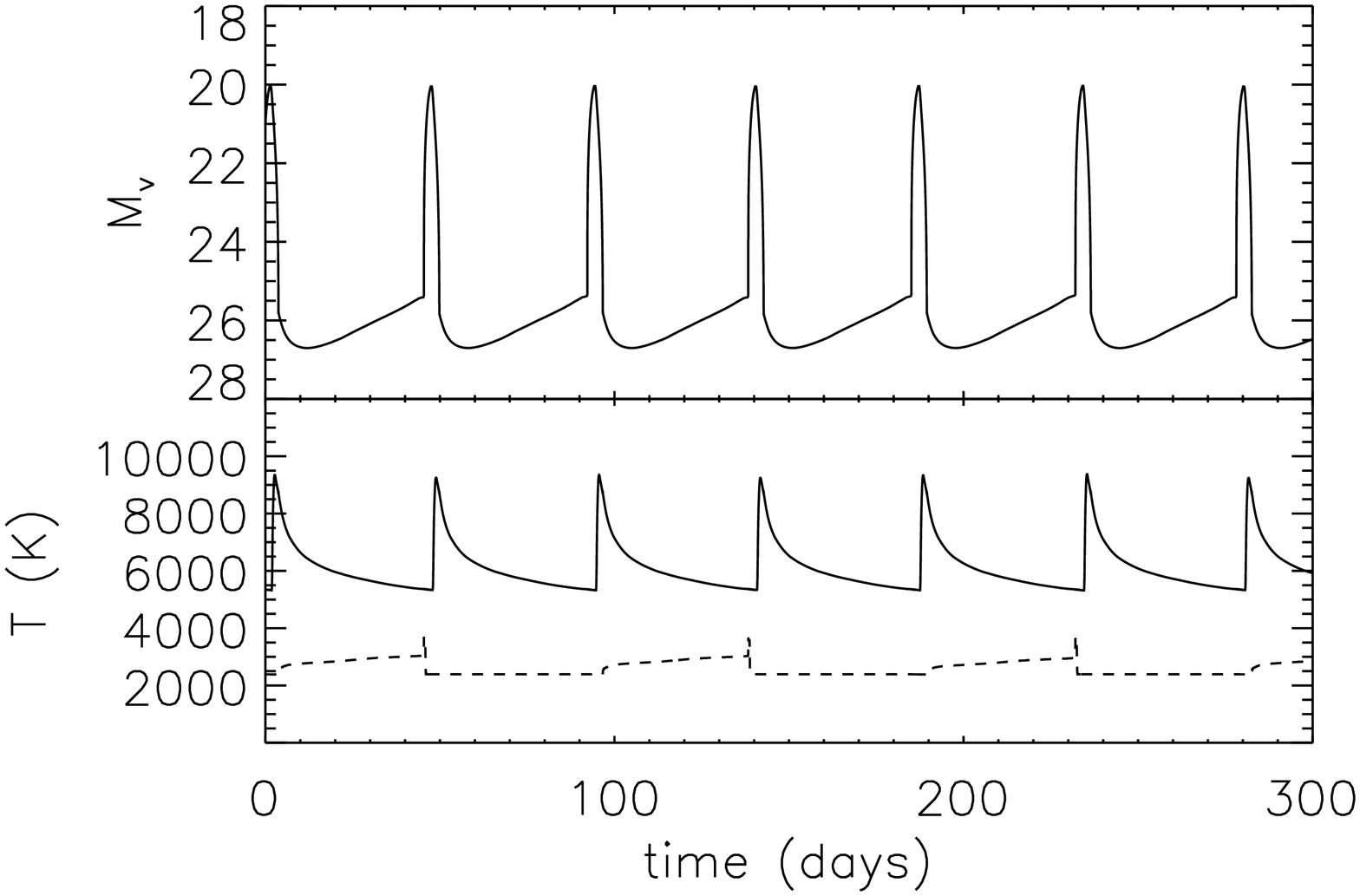}}
\parbox{0.5\linewidth}{\center \includegraphics[width=6cm,angle=90]{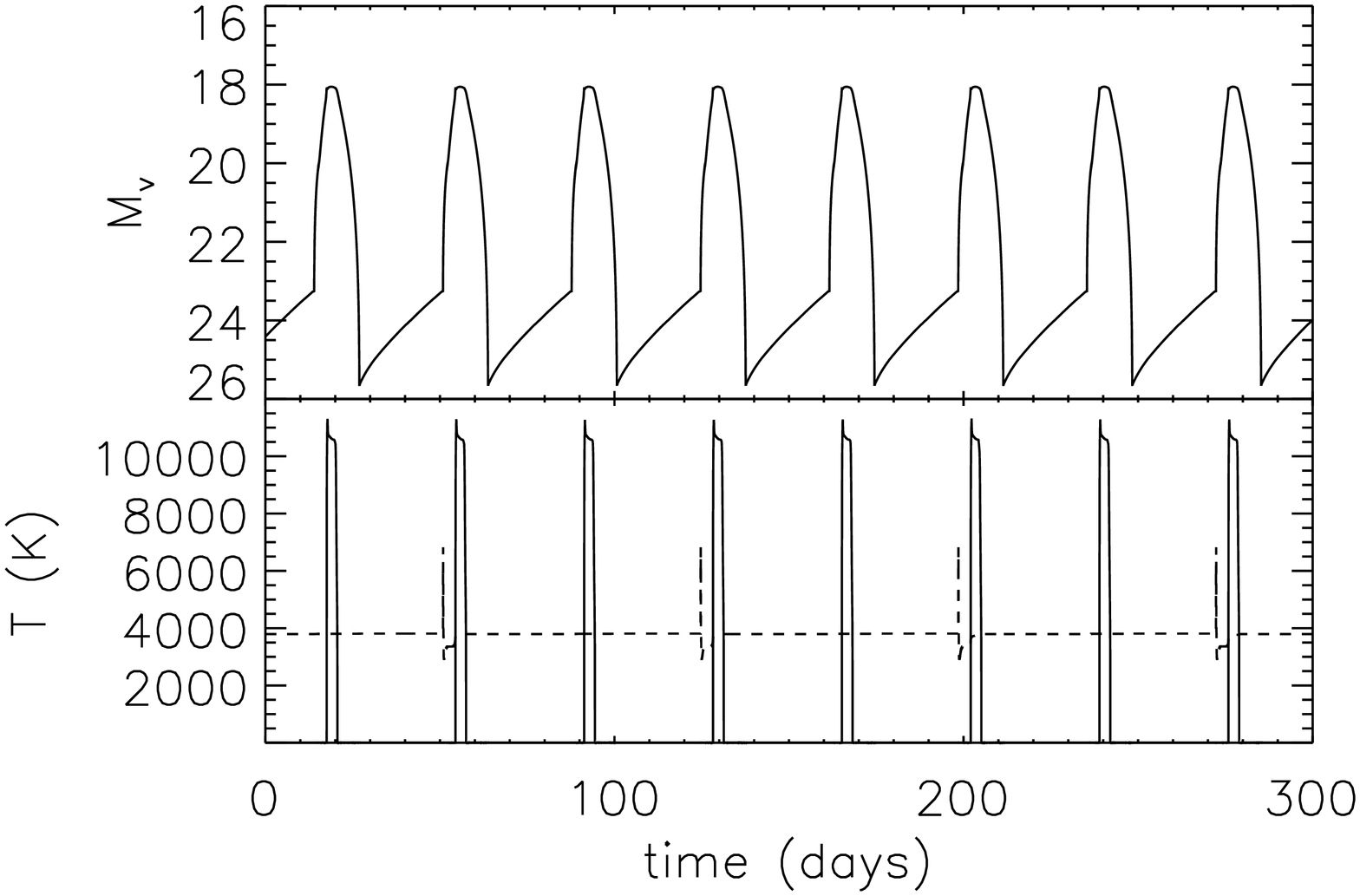}}
   \caption{Results for model 1 (left panel) and for model 2 (right panel). The top curve
   is the absolute $V$ magnitude of the system (the origin of our magnitude scale is
   arbitrary). The bottom curves show $T_\mathrm{edge}$ (plain line) and the effective
   temperature at the surface of the disc resulting from viscous dissipation (dashed line).}
   \label{oycar}
\end{figure*}

In the following, we determine whether $L_1$ can be significantly heated
by the hot edge of the disc. We consider this effect only for dwarf novae, 
since for soft X-ray transients the major effect in heating $L_1$ is due to scattering by a hot corona or an outflowing wind (see the next subsection). 
As in \cite{2002AcA....52..263S}, we first
determine the tidal torque $\dot J_\mathrm{tid}$ applied to the disc
by the secondary, equal to the rate at which angular momentum is
removed from the disc. The work done by this torque is radiated as a
``tidal luminosity" $L_\mathrm{tid} =\dot J_\mathrm{tid}\Omega$,
where $\Omega$ is the Keplerian angular velocity at the disc edge.
$L_\mathrm{tid}$ is assumed to be radiated by the disc edge, whose
effective temperature $T_\mathrm{edge}$ is then given by:

\begin{equation}
\label{Tedge}
L_\mathrm{tid} = \sigma T^4_\mathrm{edge} 2\pi R_d^2 \frac{H}{R}\Big |_{R_d}
\end{equation}

\noindent where $R_d$ is the disc outer radius and the ratio $H/R$
at the outer radius appears explicitly. $H$ is the full
geometrical disc thickness, of the order of a few times the vertical
scale height.

\cite{2002AcA....52..263S} considered the steady state case, which made the
determination of $\dot J_\mathrm{tid}$ easy. As we are interested in $\dot J_\mathrm{tid}$ during an outburst, we consider 
instead the local balance of angular momentum at the disc edge: once
the disc has reached its maximal radius, the angular momentum
transported at the disc edge has to be fully removed by the tidal
torque. Using the standard conservation laws for angular momentum
and mass, this yields:

\begin{equation}
\label{jtidal}
\dot J_\mathrm{tid} = (3\pi \nu \Sigma - \dot M_\mathrm{tr})j_d + j_s \dot M_\mathrm{tr}
\end{equation}

\noindent where $j_{\rm d}$ and $j_{\rm s}$ are respectively the
specific angular momentum at the outer disc edge and the specific
angular momentum carried by the stream. The first term on the right
hand side includes the viscous torque at the outer radius, i.e. the
rate at which angular momentum is transported outwards by viscosity,
and the effect of the incoming stream ($\dot M_\mathrm{tr}$ is the
mass transfer rate). We use the numerical code of \cite{1998MNRAS.298.1048H} 
to calculate the time evolution of the accretion disc and
thus determine $\dot J_\mathrm{tid}$. The code has been modified to
account for the different prescription for the outer disc radius.
When $R < R_\mathrm{tid}$, we use the free outer radius conditions
(Eqs. 7, 8 in \citealt{1998MNRAS.298.1048H}, note that Eq. 8 is equivalent to
$\dot J_\mathrm{tid}=0$). When $R$ reaches $R_\mathrm{tid}$, the
code switches to a fixed outer radius condition $R=R_\mathrm{tid}$.
Eq. (\ref{jtidal}) is then nonzero and yields the rate of angular
momentum removal by the tidal torque without the need for tidal terms
in the equations. The code switches back to the free outer radius
conditions when $\dot J_\mathrm{tid}$ falls below an arbitrarily
small, given value. We do not include here any heating of the outer
disc region due to the impact of the incoming stream 
\citep[see discussion in][]{2002AcA....52..263S}.

Figure 1 shows the results for models 1 (left panel) and 2 (right
panel). In our model 1, the disc always fills its tidal radius.
Outbursts are of the inside-out type and heating fronts propagate
over only $\sim 80 \%$ of the disc. The heating front then dies out
in the outer part of the disc and the corresponding outward
transport of angular momentum is large enough to counteract the
effect of the incoming stream. This contrasts with the normal tidal
torque prescription model (as described for example in \citealt{1998MNRAS.298.1048H}), 
where the outer radius of the disc is found to vary by about
$10 \%$; the reason is that the disc is too large -- despite the
short orbital period -- given the strength of the outbursts that
cannot reach the outermost parts of the disc. In Fig. 1, one can
also see that the edge of the disc is heated for a longer time than
the outburst duration. The effective temperature of disc edge is
maximum during outburst and reaches $10^4$ K. For model 2, the disc
outer radius varies, and the disc reaches its tidal radius during
outburst. The effective temperature of the edge reaches a maximum
of 1.3\ $10^4$ K during the outburst. Note that in order to obtain a
sequence of regular outbursts for this model, we used a mass
transfer rate of $10^{17}$ g.s $^{-1}$; for a value half the size,
we obtain an irregular sequence of outbursts, having the same
maximum effective temperature at the disc edge. Results for model 3 
are very similar to those of model 2.

Assuming that $L_\mathrm{tidal}$ is radiated as a blackbody, the
specific intensity of the radiation (integrated over frequency) at
the disc edge is:

\begin{equation}
\label{intensity}
I = \sigma T_\mathrm{edge}^4/\pi
\end{equation}

Let $\omega$ be the solid angle subtended by the edge of the disc at
the $L_1$ point; neglecting limb darkening and since the intensity
Eq. \ref{intensity} is isotropic, the $L_1 $ point receives an
incoming flux $F=\sigma T_\mathrm{edge}^4 (\omega/\pi)$ from the
disc edge. The computation of $\omega$ is implemented in our code
and is performed during the time evolution of the disc. The ratio
$F/F_\star$, where $F_\star = \sigma T^4_\star$ is the intrinsic
stellar flux, is given in Table 2. $F$ is equal to a few times the
intrinsic stellar flux.

\begin{table}[t]
\caption{Results for the heating of the secondary by the edge of the disc.
$T_\mathrm{edge,max}$ is the maximum effective temperature of the disc edge.
$F$ is the heating flux impinging the $L_1$ point ($F_\star$ is the intrinsic
stellar flux).}
\label{table:hotrim}      
\centering                          
\begin{tabular}{c c c c}        
\hline\hline                 
Model & $T_\mathrm{edge,max}$ (K) & $F/F_\star$\\
\hline                        
   1 & 9000 & 2\\
   2 & 13000 & 5\\
   3 & 10000& 2.5\\
\hline                         
\end{tabular}
\end{table}

As \cite{2002AcA....52..263S} emphasized, the assumption that tidal dissipation is
radiated entirely at the edge breaks down when $T_\mathrm{edge}$
becomes too different from $T_ \mathrm{surf}$, the effective
temperature at the surface of the disc due to viscous heating. In
this case, one expects an inward radial heat flux, lowering
$T_\mathrm{edge} $ and increasing $T_\mathrm{surf}$. The effective
temperature at the surface of the disc (at the outer radius) is
shown with a dashed line in Fig. 1. One can see that the ``viscous
temperature" at the outer radius remains much lower than the ``tidal
temperature". Our determination of $T_\mathrm{edge}$ is therefore
probably overestimated.

Since the secondary cannot occult the disc completely, an
enhancement of the disc edge luminosity should be detectable during
eclipses of high inclination systems. \cite{2006ApJ...644.1104S} 
analyzed eclipse profiles of the classical nova V Per. In this
system, the accretion disc is in a hot steady state and extends 
to its tidal radius. Only model profiles with a hot disc rim
($T_\mathrm{edge} \sim 6000-10000$ K) allowed the authors to find the
radial temperature profile $T \propto r^{-3/4}$ expected in the
steady state case. This range of temperature is in agreement with
the tidal heating at the outer edge. In their model, the authors
have for the first time taken into account the disc thickness which
is crucial in determining the temperature of the disc edge.
In previous studies of eclipse profiles of dwarf novae, see e.g.
\citealt{1996A&A...306..151B} (OY Car) and \citealt{1985MNRAS.214..307H} 
(Z Cha), the disc
was assumed to be ``flat" (negligible thickness), which prevented a
correct determination of the temperature at the disc edge.

\subsection{Heating by scattering of the accretion flux}

\subsubsection{The dwarf nova case: scattering by a wind}

The observation of UV lines with typical P Cyg profiles during dwarf novae outbursts implies the presence 
of mass loss in a wind \citep[see][]{2003cvs..book.....W}. Under the assumption of spherical symmetry of the wind, 
detailed studies of the line profiles have shown that the wind acceleration is slow ; the velocity profile can be taken as \citep{1987MNRAS.224..595D,1987ApJ...323..690M}:

\begin{equation}
\label{fig:wind_profile}
v(r) = v_0 + (v_\infty - v_0)(1- \frac{r_1}{r})^\beta
\end{equation}

\noindent with $\beta=6$, yielding a slow acceleration. As in \cite{1987ApJ...323..690M}, we take $v_0 = 2\times 10^{7}$ cm s$^{-1}$ 
(we will show that this parameter is of no importance in our case) and $v_\infty = 5\times 10^8$ cm s$^{-1}$, 
which is of the order of the escape velocity of the white dwarf.

Photoionisation models of the wind are consistent with a mass loss rate in the range $10^{-10}-10^{-9} M_\odot{\rm\ yr}^{-1}$, which translates into a few percent of the accretion rate \citep{1993MNRAS.260..647H}.

Due to the presence of matter above the disc, a
fraction of the accretion luminosity is scattered toward $L_1$ and
could contribute to its heating.
\begin{figure}[t] 
 \centering
\includegraphics[width=0.75\linewidth]{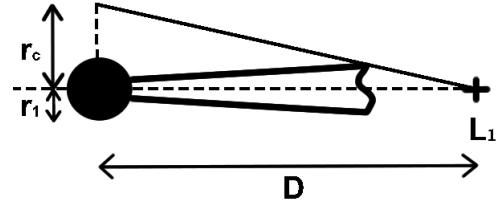}
 \caption{Side view of the binary system. The innermost part of the wind close
 to the compact object is shielded from $L_1$ by the disc and does not contribute
 to the scattered flux.}
 \label{fig:rc}
\end{figure}
We assume in the following that $5 \%$ of the accretion rate is lost
in a spherically symmetric wind. The mass conservation law yields:

\begin{equation}
4\pi r^2 \rho(r) v(r) = \alpha \dot M_\mathrm{acc,max}
\label{mass_cons}
\end{equation}

\noindent where $\dot M_\mathrm{acc,max}$ is the accretion rate onto the
compact object during an outburst (see Table 1), $\alpha=0.05$ is the $5\%$ mass
loss and $r$ is the distance to the primary center of mass. From equation \ref{mass_cons} we deduce the radial density profile of the wind.

We consider for simplicity a uniform and isotropic opacity $\kappa = 0.4$ (corresponding to
Thompson scattering by free electrons); the optical depth of the
wind is then:

\begin{equation}
\tau = \Sigma \kappa
\end{equation}

\noindent where $\Sigma$ is the total column density of the wind. 
The value of $ \tau$ for models $1-4$ is shown in Table 
\ref{table:wind}. $\tau$ is low and the wind is
optically thin. Consequently only a very small fraction of the
accretion luminosity is scattered toward $L_1$. In order to
determine this fraction, we suppose that no ``double scattering"
occurs and we neglect the radial decrease of the accretion
luminosity due to scattering losses. Scattering is handled as an
emissivity $\epsilon$ equal to:

\begin{equation}
\epsilon(r) = \frac{L_\mathrm{acc}}{16\pi^2 r^2}\rho(r)\kappa
\end{equation}

\noindent where $r$ is the distance to the primary center of mass.
The flux reflected toward $L_1$ is then:
\begin{equation}
F = \iiint \frac{\epsilon(\vec r) dV}{d^2}
\end{equation}

\noindent where $d$ is the distance between $L_1$ and the running
point $\vec r$. We show in the appendix that: 
\begin{equation}
\label{result}
F = \int_{r_c}^{\infty} \epsilon(r) g(r) dr
\end{equation}

\noindent where $g(r)$ is a geometrical weight function for the spherical shell at radius $r$. 
Note that the radial integration starts at $r_c$ (see Fig. \ref{fig:rc}), and not at $r_1$. This 
approximately takes into account the screening of the inner region of the wind by the edge of the 
accretion disc. $r_c$ is computed by considering that the disc has its maximal radius (see previous section) and typical $h/r$ ratio at the edge, see Table \ref{params_phys}. Note that due to this screening, the value $v_0$ in Eq. \ref{fig:wind_profile}, which controls the mass density of the wind near the white dwarf, does not influence the fraction that is back-scattered. 
However, it does influence the value of $\Sigma$ and hence $\tau$, so the values of $\tau$ given in Table \ref{table:wind} should be taken with care. 

Equation \ref{result} is numerically computed with the appropriate $\epsilon(r)$ profile.
In the case of a direct illumination, the $L_1$ point would receive
a flux equal to $L_\mathrm{acc}/4\pi D^2$. As shown in Table 3, the
fraction of this flux scattered toward $L_1$ is very small, of the order of $10^{-4}$. This yields a 
very low $F/F_\star$ ratio. DNe have a insufficient luminosity for this effect to be
significant.

\begin{table}
\caption{Results for the heating of the secondary by scattering of
the accretion luminosity. $\tau$ is the optical depth of the outflow
and $F$ is the heating flux impinging on the $L_1$ point.}
\label{table:wind}      
\centering                          
\begin{tabular}{c c c c c}        
\hline\hline                 
Model & $\tau$ & $F/(L_\mathrm{acc}/(4\pi D^2))$ & $F/F_\star$\\
\hline                        
   1 & 0.01 & $6\times 10^{-5}$ & $2.1\times 10^{-2}$\\
   2 & 0.1 & $6\times 10^{-5}$ & $2.1\times 10^{-2}$\\
   3 & 0.3 & $8\times 10^{-5}$ & $2.6\times 10^{-2}$\\
\hline                                   
\end{tabular}
\end{table}

\subsubsection{The X-ray transient case: scattering by a corona}

Observational evidence for an accretion disc corona in low mass X-ray 
binaries (LMXBs) was first discussed by \cite{1982ApJ...257..318W} and 
\cite{1982ApJ...258..245M}. The principal indicator of an extended region of 
hot, ionized gas above the disc is the evidence of partial eclipses of X-ray emission in 
high inclination systems. While the compact object is screened from the observer, a fraction as high as $50\%$ of the X-ray flux is left during eclipses. These sources are the so-called ``coronal sources". Such a corona is believed to be present in many, if not all, LMXBs. X-ray transients in outbursts should not be an exception.

We discuss here the possibility that, during an outburst, the $L_1$ point receives some fraction of the accretion flux due to back-scattering by the corona.

The geometry of the accretion disc corona is still poorly understood. One can grossly 
distinguish between two types of geometry: a slab corona which consists of two slabs of optically thin plasma sandwiching the accretion disc or a spherical corona located around the compact object.

Let us consider first a spherical corona. With typical X-ray transient parameters (see Table \ref{params_phys}), the value of $r_c$ (see figure \ref{fig:rc}) is $r_c \sim 10^{11}$ cm. According to studies of eclipse light-curves of coronal sources, the most extended coronae are found in the brightest sources ($L_X \sim 10^{39}$ erg s$^{-1}$) and have a radius of $7\times10^{10}$ cm \citep[see][]{2004MNRAS.348..955C}. Such a corona can be screened by the accretion disc and would not scatter radiation toward $L_1$.

A slab corona results from the X-ray heating of the accretion disc surface by the central source. The heated gas forms a tenuous layer with a thickness exceeding that of the disc. \cite{1983ApJ...271...70B} analyzed the structure and dynamics of the X-ray heated disc. They found that, in addition to the formation of the corona, X-ray heating could also drive a thermal wind in the outer region of the disc. A slab corona could back-scatter radiation coming from the central part of the disc over the disc rim, toward the $L_1$ point. A wind emitted from the outer region of the disc would also contribute to this effect. It is however hard to obtain a quantitative measure of this effect.

Results obtained in the study of irradiation of the accretion disc in X-ray transients can give us a clue about the efficiency. \cite{1999MNRAS.303..139D} show that self-consistant computation of the radial structure of irradiated discs in SXTs leads to a problem: the disc has a convex shape and the outer, unstable, regions are screened from irradiation. This is contrary to observation and the authors replace the usual formula for the irradiation temperature in the disc \citep[see e.g.][]{1973A&A....24..337S} by the following prescription:

\begin{equation}
\sigma T_\mathrm{irr}^4 = C \frac{L_{\rm acc}}{4\pi r^2}
\end{equation}

\noindent where $C$ is taken to be constant, which is thought to reflect the fact that the disc receives some X-ray flux at each radius. This could be due to warping of the disc, or to the presence of a corona. In SXT, $C=5\times 10^{-3}$ is an adequate value \citep[see][]{1999MNRAS.303..139D,2001A&A...373..251D}. Note that a warping of the disc in its outer regions could allow a direct irradiation of $L_1$. If the corona is responsible for the overall heating of the disc, $C$ could serve as a first estimate for the efficiency of the heating of the $L_1$ point by the corona. The flux impinging the $L_1$ region would therefore be:

\begin{equation}
\label{eq:cf}
F = C \frac{L_{\rm acc}}{4\pi D^2}
\end{equation}

\noindent where $D$ is the distance between $L_1$ and the compact object. SXT outbursts have typical luminosities that translate into an irradiation flux at the secondary of the order of $10^5 F_\star$. With $C=5\times 10^{-3}$, Eq. \ref{eq:cf} yields $F=500 F_\star$ which is very significant. This value is most probably overestimated ; a better determination, relying on a model for the corona and/or the wind, should be sought.

\section{Variation of $\dot M_\mathrm{tr}$}

\subsection{Vertical structure in quiescence}

The mass transfer rate is given by \citep[see][]{1975ApJ...198..383L}:

\begin{equation}
\label{mdot}
\dot M_\mathrm{tr} = Q \rho(L_1) c_s
\end{equation}

\noindent where $c_s$ is the isothermal speed of sound which depends
on the temperature at $L_1$, $T(L_1) $:
\begin{equation}
c_s = \sqrt{R_gT(L_1)}
\end{equation}

\noindent and $Q$ is the cross section of the stream \citep{1975ApJ...198..383L}:
\begin{equation}
Q=\frac{2\pi}{k}\big(\frac{c_s}{\Omega}\big)^2
\end{equation}

\noindent where $k$ depends only slightly on the mass ratio $q$.
Here we take $k = 7$, valid for $q$ in the range $0.1-0.6$ within 10 \%.

In order to determine $\rho(L_1)$ when an incident flux heats the
upper layer of the atmosphere, we first compute the structure of the
envelope of the secondary during quiescence for models $1-4$ by
solving the standard equations:

\begin{eqnarray}
\label{hydrostatic}
&&\frac{\mathrm{d}P}{\mathrm{d}r} = -\rho g(r)\\
\label{grad}
&&\frac{\mathrm{d} \ln T}{\mathrm{d} \ln P} = \nabla\\
\label{eqstate}
&&P = P(\rho, T)
\end{eqnarray}

\noindent where $\nabla$ is either the radiative temperature
gradient when $\nabla_\mathrm{rad}$ is less than the adiabatic value
$\nabla_ \mathrm{ad}$ with

\begin{equation}
\nabla_\mathrm{rad} = \frac{\kappa P F}{4 P_\mathrm{rad} c g}
\end{equation}

\noindent where $P_\mathrm{rad}$ is the radiative pressure, or the
convective value when the radiative gradient is superadiabatic
($\nabla_\mathrm {rad} > \nabla_\mathrm{ad}$), calculated in the
mixing-length prescription as described in \cite{1969AcA....19....1P}. The
equation of state and opacities are computed in the same way as in
\cite{1991A&A...243..419H}. We use the diffusion approximation for the radiative flux, which is not accurate in optically thin regions. The true radiative flux should be computed by solving the 
radiative transfer, a task which is beyond the scope of the present exploratory work. 
The photosphere of the atmosphere is defined to be the location where $\tau=2/3$, where 
$\tau$ is the Rossland optical depth.

We solve numerically Eqs. (\ref{hydrostatic}-\ref{eqstate}) assuming a given 
$T(L_1)$ and $\rho(L_1)$ at the $L_1$ point. $T(L_1)$ is chosen so 
that the temperature of the model at the photosphere is equal to the value of $T_\star$ given in Table 1.
The value of $\rho(L_1)$ is then determined from Eq. (\ref{mdot}) with the appropriate value of $T(L_1)$ and with the secular mean value of $\dot M_\mathrm{tr}$ given in Table 1. 
The computation starts at the $L_1$ point and goes inward until 
a temperature $\gtrsim 10^4$ K is reached. The Roche geometry is taken
into account by using the Roche potential in the hydrostatic
equilibrium equation.

\begin{figure}[htbp] 
   \centering
   \includegraphics[width=6.5cm,angle=90,trim=0 0 30 0]{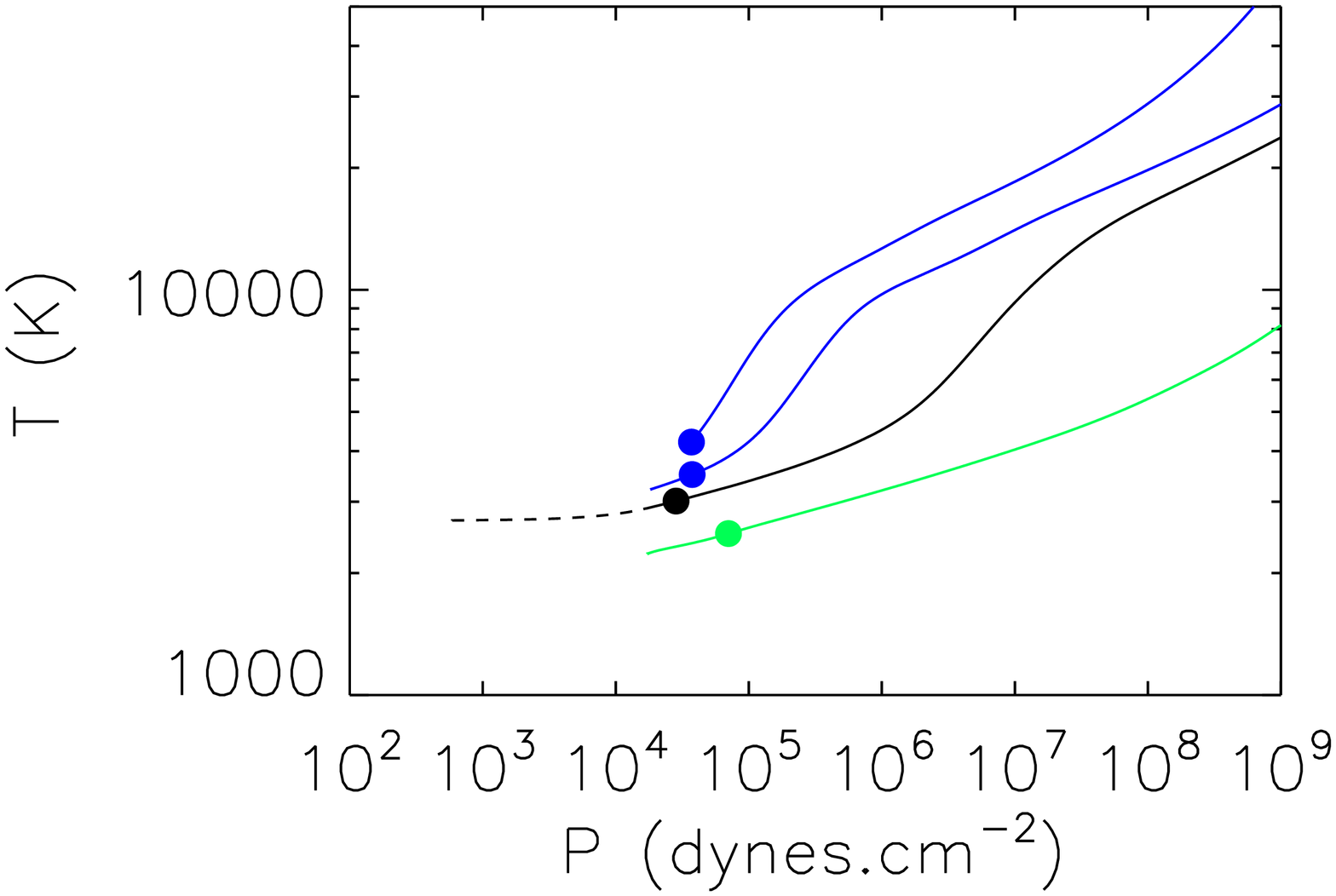}
   \caption{Vertical structure of the envelope for a secondary in DNe below the period gap
(green curve) and above the period gap (blue curves) and for a SXT (black curve). All
curves start at the $L_1$ point and end at an arbitrary depth. The line is dashed when energy transport is radiative. The dots mark the position of the photosphere, 
defined to be the location where $\tau=2/3$. The models are constructed in order to have $T(\tau=2/3)=T_\star$.}
   \label{fig:struct_vert}
\end{figure}

Figure 3 shows the vertical structure of the secondary envelope for
DNe below the period gap (model 1), DNe above the period gap (model
2 \& 3) and SXTs (model 4). As seen in Fig. \ref{fig:struct_vert}, the $L_1$ point lies in
the radiative part of the atmosphere in the SXT case and in the
convective part of the atmosphere in DNe secondaries. In each case,
the $L_1$ point is located above the photosphere, with $z_{L_1}/H
\sim 2$ in model 4,  $z_{L_1}/H \lesssim1$ in model 1\&2 and $z_{L_1}/H \sim 0.3$ 
in model 3 ($H$ is the vertical pressure scale height of the envelope).
This difference is due to the fact that SXTs have lower mass transfer rates than DNe,
which is possible since the critical mass transfer rate above which
the instability disappears is lower in an irradiated disc than in the
unirradiated case \citep[see][]{1999MNRAS.303..139D}.

Figure \ref{fig:struct_vert} also shows that for DNe below the period
gap, the temperature increases very slowly with depth in the
secondary envelope. This is due to the low value of the adiabatic
gradient ($\nabla_\mathrm{ad} \sim 0.1$) in the regions of molecular
hydrogen dissociation present in the envelope of these low
temperature stars. For DNe above the period gap, the secondary is
hotter and the $L_1$ point lies in a region where hydrogen is
neutral. As a consequence, the adiabatic gradient is larger
($\nabla_\mathrm {ad} = 0.4$) and the temperature increases rapidly
with depth.

\subsection{Vertical structure during an outburst}

We now assume that an heating flux $F_\mathrm{heat}$ strikes the 
$L_1$ region and is absorbed below the photosphere.
$F_\mathrm{heat}$ heats up the secondary on the
thermal time scale of the atmosphere, short compared to the
duration of outbursts, until the envelope reaches an outward flux
equal to:

\begin{equation}
F_\mathrm{out} = \sigma T_\star^4 + \beta F_\mathrm{heat}
\end{equation}

\noindent Here $T_\star$ is the effective
temperature of the secondary in quiescence and $\beta$ is the
bolometric albedo that accounts for a horizontal transport of the
deposited energy, which could be important in the case of deep
convective stars (for radiative stars $\beta=1$, see e.g. \citealt{1995ASPC...85..417S}). 
Since its value is not known precisely, we adopt $\beta=1$ in
all cases \citep[see][]{1993MNRAS.264..641B}. It is not clear whether on the
time scale of an outburst an efficient horizontal heat transport can
set in.

\begin{figure*}[t] 
   \centering
  \parbox{0.32\linewidth}{\includegraphics[angle=90,width=0.95\linewidth,trim=20 50 0 70]{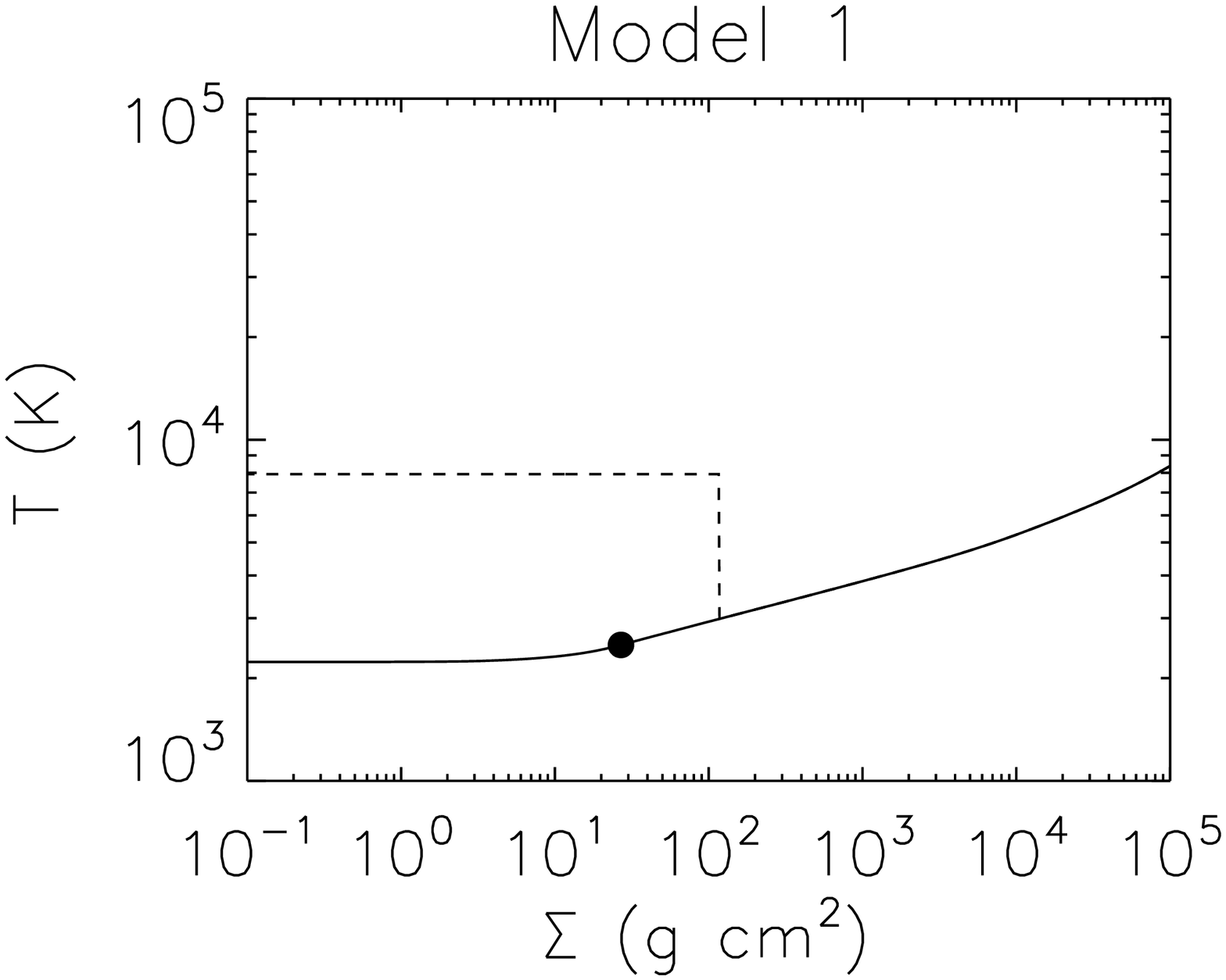}}
  \parbox{0.32\linewidth}{\includegraphics[angle=90,width=0.95\linewidth,trim=20 50 0 70]{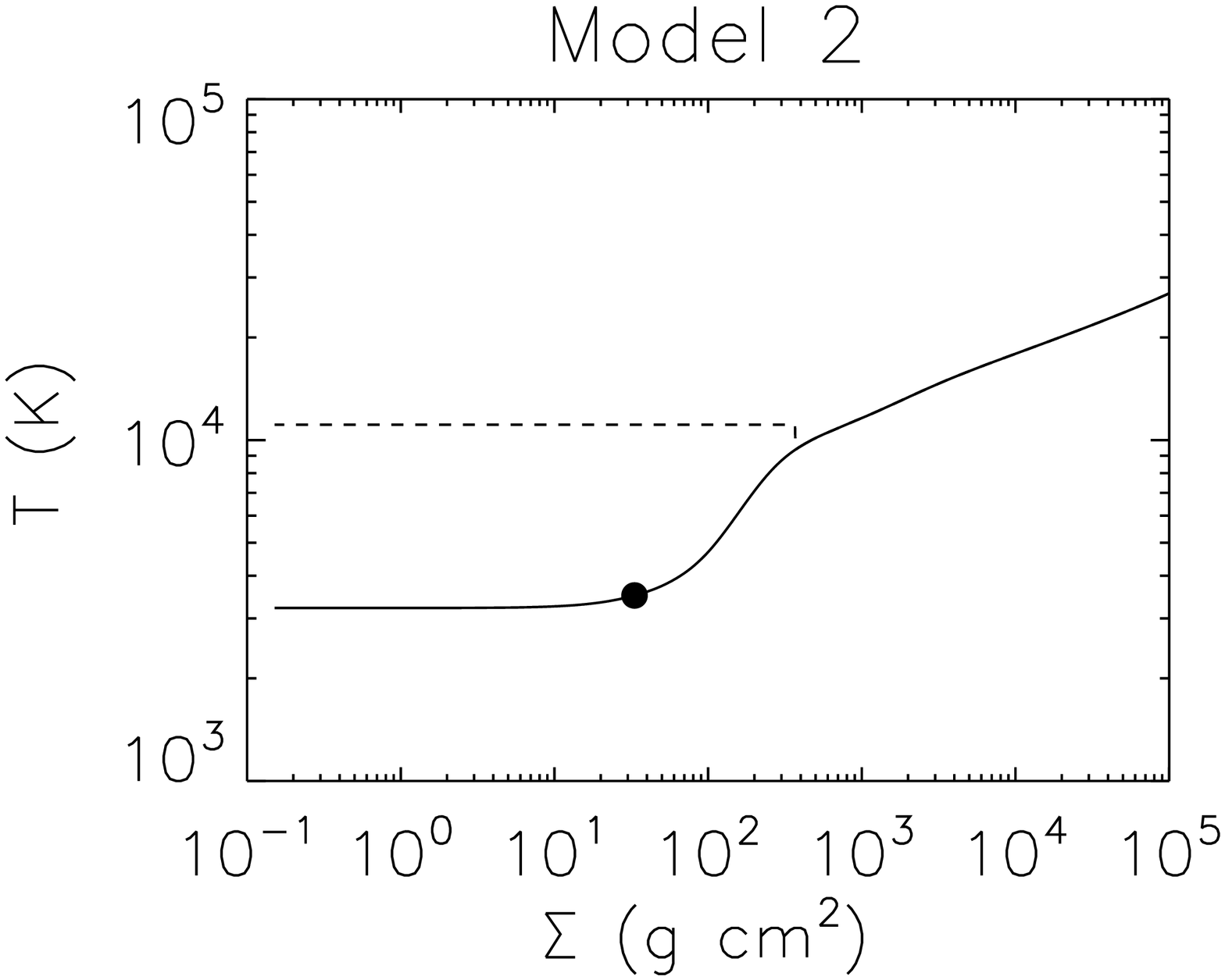}}
  \parbox{0.32\linewidth}{\includegraphics[angle=90,width=0.95\linewidth,trim=20 50 0 70]{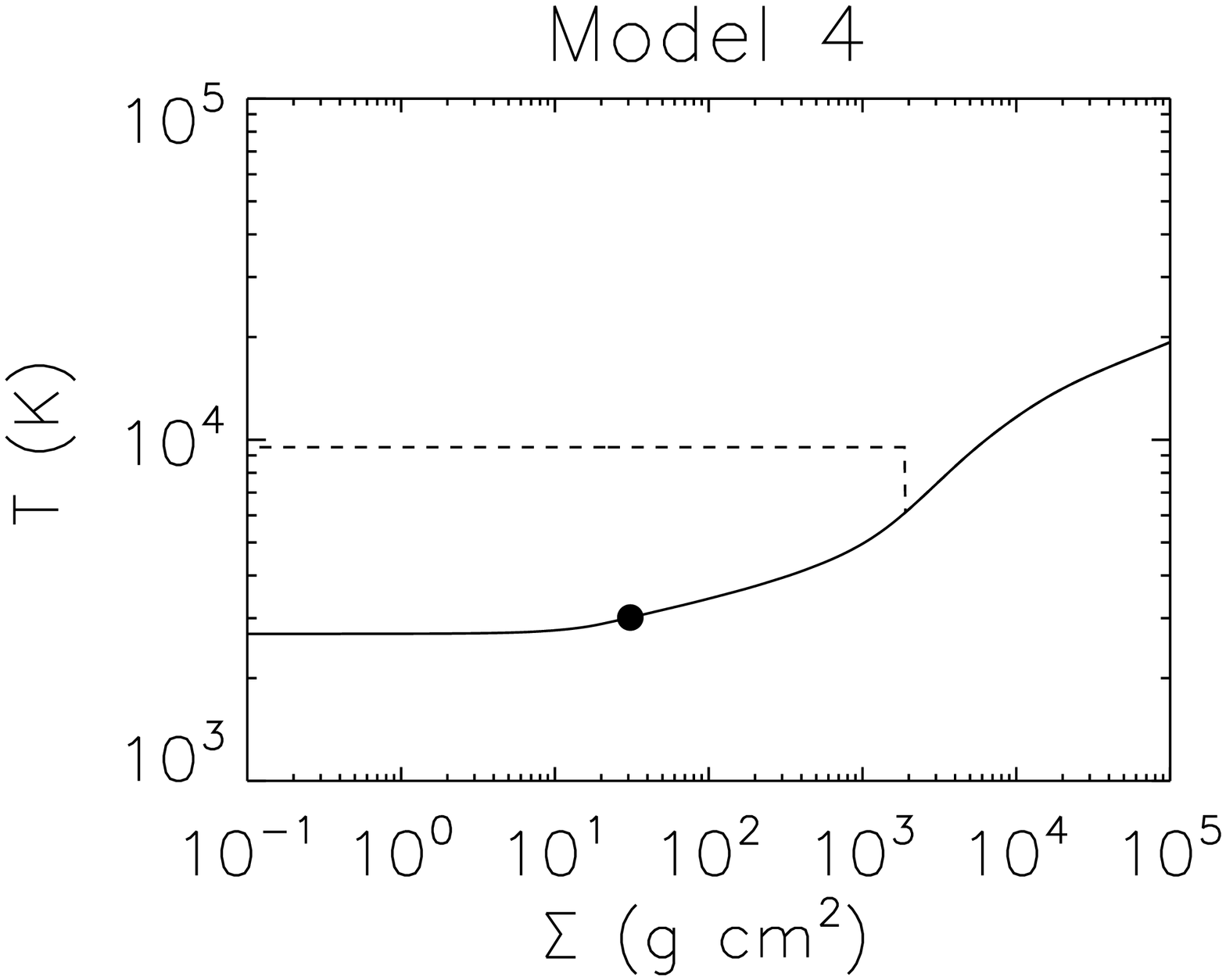}}
\caption{Temperature profiles in models of atmospheric heating by $100 F_\star$. Model 3 yields the same results than model 2.}
   \label{fig:heated_models}
\end{figure*}

The evolution of the subphotospheric layers has been determined by
\cite{1988A&A...192..187H}. As energy is deposited in sub-photospheric regions, heat diffuses inward 
and hinders the escape of the intrinsic stellar flux. An isothermal layer of temperature $T_\mathrm{layer}=(F_\mathrm{out}/\sigma)^{1/4}$ form, on top of matter that 
has not yet been heated. This heated layer ultimately extends 
to the depth where the quiescent temperature profile is equal to
$T_\mathrm{layer}$. However, assuming that the layer has been heated entirely by the stellar flux, 
its depth is restricted by the requirement that: 

\begin{equation}
\label{restriction}
E_\mathrm{th} < \Delta t_\mathrm{outburst}\sigma T^4_\star
\end{equation}

\noindent where $\Delta t$ is the duration of the outburst and
$E_\mathrm{th}$ is the thermal content of the layer; we take here
$\Delta t = 30$ $P_\mathrm{orb}$ for models $1-3$ and 
$\Delta t = 100$ $P_\mathrm{orb}$ for model $4$. We then solve Eqs.
(\ref{hydrostatic}--\ref{eqstate}) to determine the density profile of
the isothermal layer, and the new mass transfer rate is then found
from (Eq. \ref{mdot}) with the new values of $T(L_1)$ and $\rho(L_1)$.

\begin{figure}[htbp] 
   \centering
   \includegraphics[width=0.8\linewidth,angle=90]{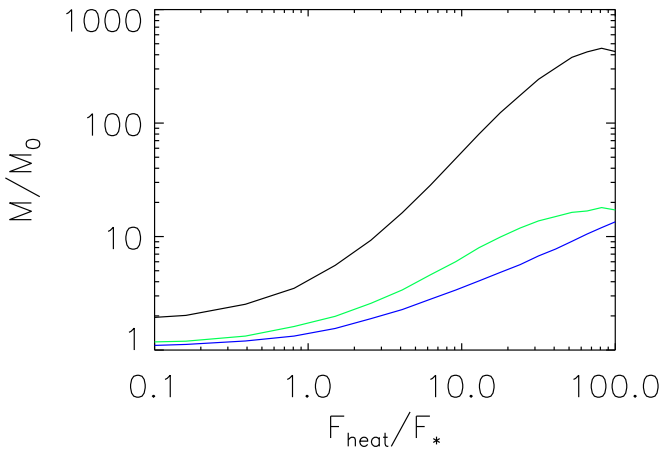}
   \caption{Mass transfer enhancement in SXT (model 4, black line), DN above the period
   gap (model 2 \& 3, blue line) and DN below the period gap (model 1, green line).}
   \label{fig:xmp}
\end{figure}

Examples of models where the heating flux is $100 F_\star$ are shown in figure \ref{fig:heated_models}. 
The figure shows that in this case, the isothermal layer extends down to a column density of the order 
of $10^2$ g cm$^{-2}$ in dwarf novae and $10^3$ g cm$^{-2}$ in soft X-ray transients. The depth of the layer is lower in dwarf novae because of the restriction (\ref{restriction}) in low period systems (SU UMa) and because of the location where $T=T_{\rm layer}$ is quickly reached in longer period systems (U Gem and Z Cam).

 
The mass transfer enhancement is computed for a heating flux 
$F_\mathrm{heat}$ ranging from $0.1F_\star$ to $100 F_\star$. 
The results are shown in Fig. \ref{fig:xmp}. As can be seen, the mass transfer rate enhancement is quite
significant in SXTs (factor of 2 for $F_\mathrm{heat} = 0.1
F_\star$) and is potentially very significant (factor of 100 for
$F_\mathrm{heat} = 10 F_\star$). In comparison, the mass transfer
enhancement is moderate in DNe. This is a direct consequence of 
the different depths reached by the isothermal layer.

\subsection{Absorption of the incident radiation by the secondary atmosphere}

The effect of an incident flux on the secondary atmosphere depends 
on the spectrum of the incoming radiation, and hence on its origin. 
The flux absorbed below the photosphere, denoted $F_\mathrm{heat}$ in the last subsection, can be quite different from the incident flux $F$.
In the case of heating of $L_1$ by the hot edge of the disc, the
spectrum of the impinging radiation can be aproximated by a
blackbody spectrum with effective temperature $\simeq 10^4$ K. 
In the case of heating by scattering of the accretion luminosity, we assume for simplicity
that no reprocessing occurs. As one half of the accretion luminosity is emitted in the accretion disc and the 
other half is emitted in the boundary layer, we use two separate contributions to the energy spectrum of the incident radiation. 
The boundary layer emission is assumed to be a black-body ; Table \ref{table:absorp} shows the wavelength of the emission
peak for each model. For the disc emission, we use a standard multi black-body spectrum of a steady disc, in which we have taken into account the irradiation by the central source in the case of SXTs. Steady state is a very reasonable assumption for discs during outburst.

\begin{table*}[t]
\caption{Fraction of the incident radiation emitted by the disc edge, by the accretion disc and by the boundary layer reaching the photospheric level ($\tau=2/3$). See text for details.}
\label{table:absorp}      
\centering                          
\begin{tabular}{c || c || c  | c | c c}        
Model & Disc edge at $T_{\rm eff} = 10^4$ K & Disc & \multicolumn{3}{c}{Boundary layer}\\
\hline
& & &$\lambda_{pic}$\ (nm) & quiescence & $F_\mathrm{heat}/F_\star = 100$ \\

   1 & 0.25 & 0.01 & 50 & $10^{-2}$ & $ 0.1$\\
   2, 3 & 0.18 & 0.05 & 30  & $ \lesssim 10^{-3}$&$ 2\ 10^{-2}$\\
   4 & 0.22 & 0.1  & 1 & $\sim 0$ & $ 0.5$\\
\hline                                   
\end{tabular}
\end{table*}

Radiation absorbed in optically thin layers contributes to the
formation of a hot corona above the photosphere ; only the incoming
radiation that penetrates below the photosphere contributes to 
changing the vertical structure and hence increasing the mass transfer rate. We use the
ATLAS 12 code to compute the monochromatic opacities in the envelope
and determine which fraction reaches the photosphere. The ATLAS 12
opacities extend from $10\ \mu$m down to 10 nm, which is a suitable
range for our purpose except for the SXT case where the bulk of the
accretion flux is emitted in the soft X-ray band (emission peak at
$\sim 1$ nm). In this last case, the opacity is extrapolated down to very short wavelengths and includes Thomson scattering. 
Both continuous and line opacities of ATLAS 12 are
described in \cite{2005MSAIS...8...25C}. Since we consider here stars with effective 
temperatures below 5000 K, molecular opacities are also
included.

\begin{figure}[t] 
   \centering
   \includegraphics[width=6cm,angle=90]{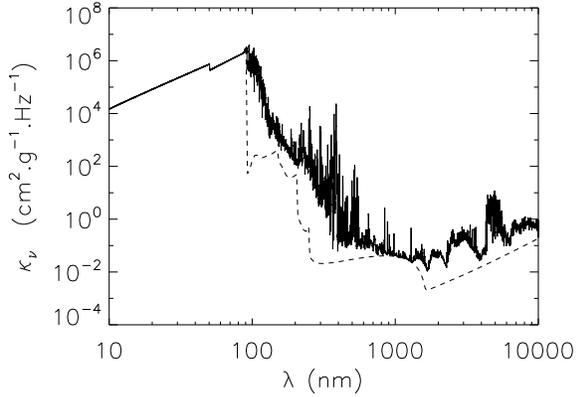}
   \caption{ATLAS 12 opacities at the photosphere of model 2, where $\rho \sim 10^{-7}$ g cm$^{-2}$ and $T= 3500$ K. The line profiles have been smoothed for readability. The dashed line is the
   continuum opacity.}
   \label{fig:kappa}
\end{figure}

Figure \ref{fig:kappa} shows the monochromatic opacity at the
photosphere of model 2. The opacity profiles of the other models show qualitatively the same behavior, the most important common feature being the very high opacity below $\sim 100$ nm due to neutral hydrogen.

We define the monochromatic optical depth $\tau_\nu$:

\begin{equation}
\tau_\nu(z) = \int_z^{z(L_1)} \rho \kappa_\nu dz
\end{equation}

\noindent We start at $L_1$ with an incident specific intensity
$B_\nu(T)$ (black-body of temperature $T$) and we compute the
specific intensity at each depth $z$ that results from absorption:
\begin{equation}
I_\nu(z) = \mathrm{e}^{-\tau_\nu(z)} B_\nu(T)
\end{equation}

Finally at the photosphere (depth $z_p$) we compute the fraction of
energy left:

\begin{equation}
\alpha(\tau_R=2/3) = \frac{\int I_\nu(z_\mathrm{p})d\nu}{\int B_\nu(T)d\nu}
= \frac{\int I_\nu(z_\mathrm{p})d\nu}{\sigma T^4/\pi}
\end{equation}

\noindent The results are shown in Table \ref{table:absorp}. For the
radiation coming from the disc edge, approximately $20-25$ \% of
the flux is able to penetrate below the photosphere. 

The radiation emitted by the boundary layer is almost completely absorbed above the 
photosphere of the secondary.
This is because the black-body maximum is located below the Lyman limit ($\lambda \sim 90$ nm) 
and absorption by neutral hydrogen is very
important. This point was already discussed by \cite{1989MNRAS.241..365K} 
in the context of cataclysmic variables. Between $1\%$ and $10\%$ of the radiation emitted by the accretion disc reaches the photosphere. 
This is because the outer region of the accretion disc emits at lower energy than the boundary layer, i.e. above $90$ nm. 
One can check that the flux left at the photosphere is mainly in the optical and near infra-red bands.

Note however that our computation here is oversimplified since we do
not consider at all the alteration of the atmosphere by the incoming
radiation.

We have also computed $\alpha$ using heated profiles we constructed in Sect 3.2. Our results show that the column density of the photosphere decreases with an increasing heating. As a consequence, an increasing fraction can be absorbed below the photosphere. Table \ref{table:absorp} shows the fraction of the incident flux absorbed in the unperturbed atmosphere of the secondary, and that absorbed in an atmosphere efficiently heated with $F_\mathrm{heat}/F_\star = 100$. We do not discuss changes for dwarf novae, since the scattered flux is far too small to have any effect. However, model 4 shows a huge difference: in quiescence nothing reaches the photospheric level whereas in the heated state, the photospheric level is so close to $L_1$ that most of the incoming flux reaches the photosphere. We therefore cannot deduce from such simple arguments whether the heating could be efficient or not. To obtain a quantitative result, one needs to solve the full radiative transfer problem. This is left for a future investigation.


\section{Conclusion}

We have considered two mechanisms that could be responsible for 
heating of the $L_1$ point during outbursts of dwarf novae and soft
X-ray transients. First, the outer edge of the disc is heated by
tidal dissipation and reaches an effective temperature of the order of
$10^4$ K (see Sect. 1). We have shown in Sect. 3 that a significant
part of the resulting thermal radiation can penetrate below the
photosphere of the secondary. The disc edge is able to heat the
upper layer of the secondary with an incoming flux comparable to the
intrinsic stellar flux. This has probably no significant effect in dwarf novae. 
In soft X-ray transients, the dominant effect could be a heating by scattered radiation. 
A disc corona, or a wind driven by the X-ray heating of the disc, could act as a scattering medium, allowing the radiation to overcome the screening by the accretion disc. It is hard however to infer a quantitative estimate of this effect. However, the luminosity of SXTs are so high that even a low efficiency ($10^{-3}-10^{-4}$) could lead to a significant heating.
However, the bulk of this radiation is absorbed by neutral hydrogen, and initially nothing reaches the photospheric level of the secondary star. Nevertheless, the flux is so strong that the incident radiation probably strongly modifies the state of the atmosphere. Yet it is hard to assess whether in this case 
an efficient heating takes place or not. The resolution of the corresponding radiative 
transfer should be undertaken.

In Sect. 3 we determined the relation between the mass
transfer enhancement and a given heating flux impinging the $L_1$
point (independently of the origin of this heating). We computed
the vertical structure of the secondary with the assumption of
hydrostatic equilibrium. The atmosphere is not hydrostatic near the $L_1$ point due to the mass outflow in
this region. However, departure from hydrostatic equilibrium is
likely to be important only in the near vicinity of the $L_1$ point.
We found that the $L_1$ point lies higher in the atmosphere in soft X-ray transients
than in dwarf novae. This is a consequence of a lower value of the mass
transfer rate in soft X-ray transients, where the $L_1$ point lies in a nearly
isothermal, radiative region of the atmosphere, whereas in dwarf novae the
$L_1$ point lies in the convective part of the atmosphere, where
temperature increases more rapidly with depth. As a consequence, an
incident flux can heat a significant layer in soft X-ray transients, which yields a large
enhancement of the mass transfer rate. On the other hand, in dwarf novae only
a narrower region of the atmosphere is affected. For a dwarf nova below the
period gap, the envelope is too massive to be fully affected on the
time scale of an outburst. Our results suggest that the mass transfer
enhancement could be potentially very important for soft X-ray
transients, up to a factor of $\sim 100$ for a heating flux equal to
$10 F_\star$. For dwarf novae, the mass transfer rate enhancement is
more moderate, of a factor of $\sim 10$ for a heating flux equal to
$10 F_\star$. In dwarf novae, none of the effects investigated here is able to
produce such fluxes. The soft X-ray transient case needs further investigation, on one hand 
to determine the fraction of radiation that is back-scattered toward $L_1$, and on the other hand the efficiency of the heating of the secondary has to be determined properly.

\begin{acknowledgements}
We gratefully thank Irit Idan for her help with the ATLAS 12 opacities. We thank the anonymous referee for very useful comments that helped us to improve the paper.
\end{acknowledgements}

\bibliographystyle{aa}
\bibliography{biblio}

\begin{thebibliography}{41}
\expandafter\ifx\csname natexlab\endcsname\relax\def\natexlab#1{#1}\fi

\bibitem[{{Augusteijn} {et~al.}(1993){Augusteijn}, {Kuulkers}, \&
  {Shaham}}]{1993A&A...279L..13A}
{Augusteijn}, T., {Kuulkers}, E., \& {Shaham}, J. 1993, \aap, 279, L13

\bibitem[{{Begelman} {et~al.}(1983){Begelman}, {McKee}, \&
  {Shields}}]{1983ApJ...271...70B}
{Begelman}, M.~C., {McKee}, C.~F., \& {Shields}, G.~A. 1983, \apj, 271, 70

\bibitem[{{Brett} \& {Smith}(1993)}]{1993MNRAS.264..641B}
{Brett}, J.~M. \& {Smith}, R.~C. 1993, \mnras, 264, 641

\bibitem[{{Bruch} {et~al.}(1996){Bruch}, {Beele}, \&
  {Baptista}}]{1996A&A...306..151B}
{Bruch}, A., {Beele}, D., \& {Baptista}, R. 1996, \aap, 306, 151

\bibitem[{{Buat-M{\'e}nard} {et~al.}(2001){Buat-M{\'e}nard}, {Hameury}, \&
  {Lasota}}]{2001A&A...366..612B}
{Buat-M{\'e}nard}, V., {Hameury}, J.-M., \& {Lasota}, J.-P. 2001, \aap, 366,
  612

\bibitem[{{Castelli}(2005)}]{2005MSAIS...8...25C}
{Castelli}, F. 2005, Memorie della Societa Astronomica Italiana Supplement, 8,
  25

\bibitem[{{Chen} {et~al.}(1993){Chen}, {Livio}, \&
  {Gehrels}}]{1993ApJ...408L...5C}
{Chen}, W., {Livio}, M., \& {Gehrels}, N. 1993, \apjl, 408, L5

\bibitem[{{Church} \& {Ba{\l}uci{\'n}ska-Church}(2004)}]{2004MNRAS.348..955C}
{Church}, M.~J. \& {Ba{\l}uci{\'n}ska-Church}, M. 2004, \mnras, 348, 955

\bibitem[{{Drew}(1987)}]{1987MNRAS.224..595D}
{Drew}, J.~E. 1987, \mnras, 224, 595

\bibitem[{{Dubus} {et~al.}(2001){Dubus}, {Hameury}, \&
  {Lasota}}]{2001A&A...373..251D}
{Dubus}, G., {Hameury}, J.-M., \& {Lasota}, J.-P. 2001, \aap, 373, 251

\bibitem[{{Dubus} {et~al.}(1999){Dubus}, {Lasota}, {Hameury}, \&
  {Charles}}]{1999MNRAS.303..139D}
{Dubus}, G., {Lasota}, J.-P., {Hameury}, J.-M., \& {Charles}, P. 1999, \mnras,
  303, 139

\bibitem[{{Hameury}(1991)}]{1991A&A...243..419H}
{Hameury}, J.~M. 1991, \aap, 243, 419

\bibitem[{{Hameury}(2000)}]{2000NewAR..44...15H}
{Hameury}, J.-M. 2000, New Astronomy Review, 44, 15

\bibitem[{{Hameury} {et~al.}(1986){Hameury}, {King}, \&
  {Lasota}}]{1986A&A...162...71H}
{Hameury}, J.~M., {King}, A.~R., \& {Lasota}, J.~P. 1986, \aap, 162, 71

\bibitem[{{Hameury} \& {Lasota}(2005)}]{2005A&A...443..283H}
{Hameury}, J.-M. \& {Lasota}, J.-P. 2005, \aap, 443, 283

\bibitem[{{Hameury} {et~al.}(1988){Hameury}, {Lasota}, \&
  {King}}]{1988A&A...192..187H}
{Hameury}, J.~M., {Lasota}, J.~P., \& {King}, A.~R. 1988, \aap, 192, 187

\bibitem[{{Hameury} {et~al.}(1998){Hameury}, {Menou}, {Dubus}, {Lasota}, \&
  {Hure}}]{1998MNRAS.298.1048H}
{Hameury}, J.-M., {Menou}, K., {Dubus}, G., {Lasota}, J.-P., \& {Hure}, J.-M.
  1998, \mnras, 298, 1048

\bibitem[{{Hoare} \& {Drew}(1993)}]{1993MNRAS.260..647H}
{Hoare}, M.~G. \& {Drew}, J.~E. 1993, \mnras, 260, 647

\bibitem[{{Horne} \& {Cook}(1985)}]{1985MNRAS.214..307H}
{Horne}, K. \& {Cook}, M.~C. 1985, \mnras, 214, 307

\bibitem[{{Ichikawa} \& {Osaki}(1994)}]{1994PASJ...46..621I}
{Ichikawa}, S. \& {Osaki}, Y. 1994, \pasj, 46, 621

\bibitem[{{King}(1989)}]{1989MNRAS.241..365K}
{King}, A.~R. 1989, \mnras, 241, 365

\bibitem[{{King} \& {Ritter}(1998)}]{1998MNRAS.293L..42K}
{King}, A.~R. \& {Ritter}, H. 1998, \mnras, 293, L42

\bibitem[{{Lasota}(2001)}]{2001NewAR..45..449L}
{Lasota}, J.-P. 2001, New Astronomy Review, 45, 449

\bibitem[{{Lewin} {et~al.}(1997){Lewin}, {van Paradijs}, \& {van den
  Heuvel}}]{1997xrb..book.....L}
{Lewin}, W.~H.~G., {van Paradijs}, J., \& {van den Heuvel}, E.~P.~J. 1997,
  {X-ray Binaries} (X-ray Binaries, Edited by Walter H.~G.~Lewin and Jan van
  Paradijs and Edward P.~J.~van den Heuvel, pp.~674.~ISBN
  0521599342.~Cambridge, UK: Cambridge University Press, January 1997.)

\bibitem[{{Lubow} \& {Shu}(1975)}]{1975ApJ...198..383L}
{Lubow}, S.~H. \& {Shu}, F.~H. 1975, \apj, 198, 383

\bibitem[{{Mauche} \& {Raymond}(1987)}]{1987ApJ...323..690M}
{Mauche}, C.~W. \& {Raymond}, J.~C. 1987, \apj, 323, 690

\bibitem[{{McClintock} {et~al.}(1982){McClintock}, {London}, {Bond}, \&
  {Grauer}}]{1982ApJ...258..245M}
{McClintock}, J.~E., {London}, R.~A., {Bond}, H.~E., \& {Grauer}, A.~D. 1982,
  \apj, 258, 245

\bibitem[{{Osaki}(1989)}]{1989PASJ...41.1005O}
{Osaki}, Y. 1989, \pasj, 41, 1005

\bibitem[{{Osaki}(1996)}]{1996PASP..108...39O}
{Osaki}, Y. 1996, \pasp, 108, 39

\bibitem[{{Paczy{\'n}ski}(1969)}]{1969AcA....19....1P}
{Paczy{\'n}ski}, B. 1969, Acta Astronomica, 19, 1

\bibitem[{{Shafter} \& {Misselt}(2006)}]{2006ApJ...644.1104S}
{Shafter}, A.~W. \& {Misselt}, K.~A. 2006, \apj, 644, 1104

\bibitem[{{Shakura} \& {Syunyaev}(1973)}]{1973A&A....24..337S}
{Shakura}, N.~I. \& {Syunyaev}, R.~A. 1973, \aap, 24, 337

\bibitem[{{Smak}(1999)}]{1999AcA....49..383S}
{Smak}, J. 1999, Acta Astronomica, 49, 383

\bibitem[{{Smak}(2002)}]{2002AcA....52..263S}
{Smak}, J. 2002, Acta Astronomica, 52, 263

\bibitem[{{Smak}(2004)}]{2004AcA....54..181S}
{Smak}, J. 2004, Acta Astronomica, 54, 181

\bibitem[{{Smith}(1995)}]{1995ASPC...85..417S}
{Smith}, B.~C. 1995, in Astronomical Society of the Pacific Conference Series,
  Vol.~85, Magnetic Cataclysmic Variables, ed. D.~A.~H. {Buckley} \&
  B.~{Warner}, 417--+

\bibitem[{{Truss} {et~al.}(2002){Truss}, {Wynn}, {Murray}, \&
  {King}}]{2002MNRAS.337.1329T}
{Truss}, M.~R., {Wynn}, G.~A., {Murray}, J.~R., \& {King}, A.~R. 2002, \mnras,
  337, 1329

\bibitem[{{Viallet} \& {Hameury}(2007)}]{2007A&A...475..597V}
{Viallet}, M. \& {Hameury}, J.-M. 2007, \aap, 475, 597

\bibitem[{{Warner}(2003)}]{2003cvs..book.....W}
{Warner}, B. 2003, {Cataclysmic Variable Stars} (Cataclysmic Variable Stars, by
  Brian Warner, pp.~592.~ISBN 052154209X.~Cambridge, UK: Cambridge University
  Press, September 2003.)

\bibitem[{{White} \& {Holt}(1982)}]{1982ApJ...257..318W}
{White}, N.~E. \& {Holt}, S.~S. 1982, \apj, 257, 318

\bibitem[{{Whitehurst}(1988)}]{1988MNRAS.232...35W}
{Whitehurst}, R. 1988, \mnras, 232, 35

\end{thebibliography}

\appendix

\section{Derivation of Eq. \ref{result}}

We give here the detailed computation of the integral:

\begin{equation}
\label{inte}
F = \iiint \frac{\epsilon(\vec r) dV}{d^2}
\end{equation}

\noindent with
\begin{equation}
\epsilon(r) = \frac{L_\mathrm{acc}}{16\pi^2 r^2}\rho(r)\kappa \mathrm{\ \ \ and\ \ \ } \rho
(r) = \rho_1\big(\frac{r_1}{r}\big)^{3/2}
\end{equation}

We adopt a spherical system of coordinates centered on the primary
center of mass, with the colatitude coordinate $\theta$ defined to be
the angle between the running point $\vec r$ and the line joining
the primary center of mass to $L_1$ and with longitude coordinate
$\phi$. Due to the symmetry of the problem, the integral
\ref{inte} becomes:

\begin{eqnarray}
\label{int2}
F &=& \int_{r_c}^\infty dr \epsilon(r) \iint \frac{dS}{d^2} \nonumber \\ 
   &=& \int_{r_c}^\infty dr \epsilon(r) g(r)
\end{eqnarray}
\noindent with $dS=r^2 \sin \theta d \theta d \phi$ and $d^2 = r^2 + D^2 - 2rD\cos \theta$.
Note that the radial integration starts at $r=r_c$ (see Fig. \ref{fig:rc}) and not $r=r_1$ this
approximately accounts for the shielding of the scattering region by the disc rim. The double 
integral involving the angular variables gives:

\begin{eqnarray}
\label{w}
g(r) = \iint \frac{dS}{d^2} = 2\pi \frac{r}{D} \ln \frac{1+ \frac{D}{r}}{|1- \frac{D}{r}|}
\end{eqnarray}

It is useful to look at the asymptotic behavior
of this function: one can check that $g(r)\rightarrow 4\pi$ as $r \rightarrow \infty$ 
and that $g(r) \rightarrow 0$ as $r \rightarrow 0$. Note that this
quantity diverges at $r=D$, however the integral Eq. \ref{int2} is
well defined because the divergence of the logarithm function at the 
origin is integrable.

\end{document}